\documentclass[
 reprint,
 amsmath,amssymb,
 aps,
]{revtex4-1}

\usepackage{graphicx}
\usepackage{dcolumn}
\usepackage{bm}
\usepackage{subcaption}

\begin{document}

\preprint{APS/123-QED}

\title{The optical performance of a dielectric-metal-dielectric anti-reflective absorber structure}

\author{V. V. Medvedev}
\altaffiliation[Also at ]{Moscow Institute of Physics and Technology (State University), Institutskiy pereulok str. 9, Dolgoprudny, Moscow region 141701, Russia}
\email{medvedev@phystech.edu}
\affiliation{Institute for Spectroscopy RAS, Fizicheskaya str. 5, Troitsk, Moscow 108840, Russia}

\author{V. M. Gubarev}
\affiliation{Moscow Institute of Physics and Technology (State University), Institutskiy pereulok str. 9, Dolgoprudny, Moscow region 141701, Russia}
\author{C. J. Lee}
\affiliation{Insitute of Engineering, Hogescholen Fontys, Eindhoven, The Netherlands}

\date{\today}

\begin{abstract}
The absorption of electromagnetic radiation by a planar structure, consisting of a three-layer dielectric-metal-dielectric coating on a metal back-reflector is analysed. The conditions for total absorption are derived. Our analysis shows that, in contrast with bi-layer structures, the calculated layer thicknesses are feasible to fabricate for any metal. The proposed absorber design is of potential use in infrared, terahertz and longer wavelength detectors and for radiant energy harvesting devices.
\end{abstract}

\pacs{Valid PACS appear here}
\maketitle


\section{Introduction}
Metal films with nanometer-scaled thicknesses can effectively absorb electromagnetic radiation \cite{Schossig2012}. Fundamental aspects of this have been studied up to recent years \cite{li2015cpa,luo2014unified,li2014cpaSciRep}. Ultra-thin metal films with nanometer-scaled thicknesses also serve as important components in high-efficiency absorber structures for electromagnetic radiation. 
These absorber structures are widely used in uncooled detectors for infrared, terahertz and longer wavelength radiation. Examples of such detectors are resistance bolometers, pyroelectric and ferroelectric detectors \cite{Parsons:88,Bauer1992,Bly:94,Thompson:2007,talghader2012,Jung:14,Demyanenko:17}. Absorbers utilizing thin metal films are also attractive for applications in radiant energy harvesting from remote or nearby heat sources. These can be absorbers for near-infrared solar radiant energy \cite{Sergeant:09,Chirumamilla:16,Chen:16} or infrared radiated from engines or factories as waste heat. Such harvesters have the potential to turn waste heat into usable energy \cite{Corrigan:12,Guo:14}. Besides that, infrared absorbing structures can also be used for thermal shielding and camouflaging \cite{Peng:16}. The absorptive properties of thin metal films can also be used to design complex spectral filters \cite{Dobrowolski:95,medvedev2011infrared}. Finally, making use of the time-reverse symmetry of absorption and emission, thin metal films can also be used as high efficiency quasi-monochromatic infrared emitters \cite{brucoli:14}.

\begin{figure}[htbp]
\centering
\includegraphics[width=\linewidth]{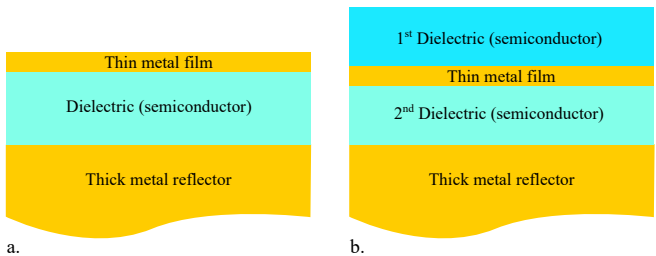}
\caption{a) sketch of a two layer coating (thin metal film and lossless layer) on an optically thick metal substrate; b) sketch of a three layer coating (lossless layer, thin metal film and lossless layer) on an optically thick metal substrate.}
\label{fig:figure1}
\end{figure}

An absorber structure of fundamental importance for the above-mentioned applications is the so-called quarter-wavelength absorber (QWA) or Salisbury screen \cite{BauerAmJPhys1992,fante1988}. The QWA structure is shown schematically  in Fig. \ref{fig:figure1}a. It consists of a thin metal film and a metal mirror-like substrate that are separated by a spacer layer of a transparent (lossless) material, e.g. dielectric or lossless semiconductor. The theory of absorption of electromagnetic radiation by such structures has been well studied \cite{Hadley:47,Hilsum:54,Silberg:57,fante1988,Monzon:94,Razansky:06}. Theoretical analysis shows that the QWA structure  achieves total absorption at a given wavelength, $\lambda$, when the following conditions are satisfied \cite{fante1988,BauerAmJPhys1992}:
\begin{equation}
R_{sh} = Z_0
\label{eq1}
\end{equation}
\begin{equation}
d_{D} = \lambda (2m-1)/4n_D
\label{eq2}
\end{equation}
In Eq. (\ref{eq1}), $R_{sh}$ denotes the  electrical sheet resistance of the top absorbing metal layer and $Z_0 = \sqrt[]{\mu_0/\varepsilon_0} \approx 377~ \Omega$ denotes the vacuum impedance. The sheet resistance is defined as $R_{sh} = 1/\sigma d$, where $d$ and $\sigma$ stand for the thickness and volumetric electrical conductivity of the top absorbing layer. In Eq. (\ref{eq2}), $d_D$ and $n_D$ denote thickness and refractive index of the dielectric spacer layer and $m = 1,2,3,...$ QWA-based structures are usually fabricated with the minimal thickness of the dielectric layer, i.e. with $d_{D} = \lambda/4n_D$. Eq. (\ref{eq1}) allows the optimal layer thickness of the absorber layer to be estimated. Using the tabulated values of electrical conductivity of metals one can easily find that, for highly conductive metals (e.g. Cu, Au or Al) and for moderately conductive metals (e.g. Ni, Mo or W), the optimal layer thicknesses are of sub-nanometer scale. Thus, the conditions of total absorption pose significant limitations on the absorber material choice. For that reason low conductivity metals, such as Ti and Cr or even alloys like NiCr, are used to fabricate QWA-based structures \cite{Parsons:88,Jung:14,biener2007}.

In this paper, we present the analysis of the absorption of electromagnetic radiation by a modified structure, shown in Fig. \ref{fig:figure1}b. In the considered structure, a thin metal film is sandwiched between two transparent (lossless) layers and this three layer system is placed onto an optically thick metal substrate. We denote thicknesses and refractive indices of the top and bottom dielectric layers as $d_1$, $n_1$ and $d_2$, $n_2$, respectively. An analytical theory of the interaction of a plane electromagnetic wave with the structure is presented. The conditions of total absorption at a given wavelength are derived from the analytical theory. We show that this minimal modification of the structure--the addition of a dielectric top layer--drastically changes the conditions of total absorption. We show that total absorption at a given wavelength can be achieved for an arbitrary value of the thickness of the upper layer. In this case, by varying the thickness and refractive index of the upper layer, it is possible to vary the optimum thicknesses of the other two layers. And most importantly, we show that the maximal thickness of the absorbing metal layer for the proposed structure exceeds that for QWA by a factor of $n_{1}^{2}$. This is achieved when $d_1 = \lambda (2k-1)/4n_1$ with $k = 1,2,3,...$ and $d_2 = \lambda (2m-1)/4n_2$ with $m = 1,2,3,...$ Thus, by using high-refractive materials such as Si or Ge, it becomes possible to bring the absorber layer thickness to the nanometer scale, which is feasible for state-of-art thin-film fabrication methods. Besides that, an extra dielectric layer, placed on top of the thin metal film, might improve other functional characteristics of the structure. For instance, the top dielectric layer can also serve as a protective cap, improving lifetime of the absorber structure.

\section{Theory}

We start by discussing the model assumptions and approximations that are used in the theoretical analysis. First, we assume that the optical properties of a thin metal film are determined by its electrical conductivity, $\sigma$, and thickness, $d$. This assumption is valid for infrared and longer wavelength radiation. For a given wavelength, $\lambda$, of the incident radiation, the interference-enhanced absorption in the metal film requires its thickness to be much smaller than the skin depth of the metal. Under these conditions, the propagation of the electromagnetic fields through the metal layer can be neglected. Such an ultra-thin layer can be approximated by a conducting sheet of infinitesimal thickness. The sheet electrical conductivity, $\sigma_{s}$, is related to the conductivity and the thickness of the initial metal film via $\sigma_{s} = \sigma d$. Hence, the three-layer coating shown in Fig. 1B is modeled by the two-layer coating composed of two lossless (insulating) layers that are separated by a conducting  interface. The thicknesses and refractive indices of these two layers are denoted as $d_1$, $d_2$ and $n_1$, $n_2$. Finally, the metal reflector situated at the bottom of the absorber structure shown in Fig. 1B is approximated by a perfect electrical conductor (PEC). This means that the electromagnetic wave propagating through the layered structure does not penetrate into the reflector and the electric field collapses at the reflector’s surface.

\begin{figure}[htbp]
\centering
\includegraphics[width=0.8 \linewidth]{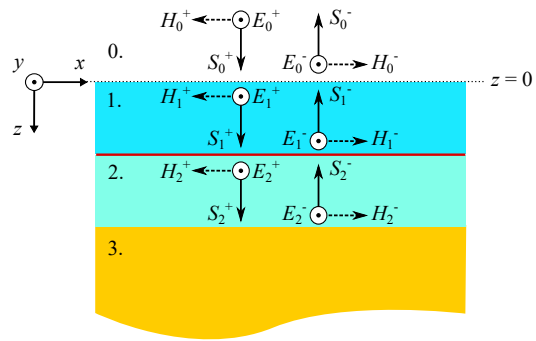}
\caption{The model structure composed of four zones: the 0th zone is ambient with refractive index $n_0$; the 1st zone is the top lossless dielectric layer with thickness $d_1$ and refractive index $n_1$; the 2nd zone is the bottom lossless dielectric layer with thickness $d_2$ and refractive index $n_2$; the red line denotes the conducting interface that models an ultra-thin metal film; the 3rd zone is a perfect electrical conductor. The reference frame is shown on the left; $y$ axis is directed out of the page. In 0th, 1st, and 2nd zones, dashed arrows denote magnetic field vectors, and solid arrows denote Poynting vectors.}
\label{fig:figure2}
\end{figure}

Let us consider the interaction of a plane electromagnetic wave with the model structure described above - see Fig. \ref{fig:figure2}. We limit ourselves to the case of a normally incident wave, which is the most relevant case for applications. For each layer of the structure and for the half-space above the structure, the electric and magnetic field components can be written as:
\begin{equation}
E_{j,y} = E_j^{+}e^{i k_{j} z} + E_j^{-}e^{-i k_{j} z},
\label{eq3}
\end{equation}
\begin{equation}
H_{j,x} = -\varepsilon_0 c n_j \left(E_j^{+}e^{i k_{j} z} - E_j^{-}e^{-i k_{j} z}\right),
\label{eq4}
\end{equation}
where the subscript $j$ denotes the zone number according to Fig. \ref{fig:figure2}, $k_{j}$ ($k_{j}= n_j (\omega/c) = n_j (2\pi/\lambda)$) is the wave vector in the $j$-th zone, $n_j$ is the refractive index the material in the $j$-th zone, $\omega$ is the angular frequency, $c$ is the speed of light in vacuum, $\varepsilon_0$ is the permittivity of vacuum. The plus (minus) superscripts indicate a wave traveling into (out of) the structure. For a given amplitude $E_{0}^{+}$  of the incident wave, one can find all amplitude coefficients $E_{j}^{+}$ and $E_{j}^{-}$ by substituting Eq (\ref{eq3}) and (\ref{eq4}) into the boundary conditions for the electric and magnetic fields for the zone interfaces. For the interface between the 0th and 1st zones, boundary conditions, in the form of the continuity of the tangential field components $E_{j,y}$ and $H_{j,x}$ give:
\begin{equation}
E_0^{+} + E_0^{-} = E_1^{+} + E_1^{-},
\label{eq5}
\end{equation}
\begin{equation}
n_0 (E_0^{+} - E_0^{-}) = n_1 (E_1^{+} - E_1^{-}).
\label{eq6}
\end{equation}
For the interface between the 1st and 2nd zones, the continuity of the tangential component of the electric field remains valid and gives:
\begin{equation}
E_1^{+} e^{i k_{1} d_1} + E_1^{-} e^{-i k_{1} d_1} = E_2^{+} e^{i k_{2} d_1} + E_2^{-} e^{-i k_{2} d_1}.
\label{eq7}
\end{equation}
But, the tangential component of the magnetic field changes at this interface according to $H_{2,x} - H_{1,x} = j_{s,y}$. $j_{s,y}$ is the component of the surface electric current density along the $y$ axis. Using the modified boundary condition for the magnetic field and Ohm's law ($j_{s}=\sigma_s E$) one can write:
\begin{widetext}
\begin{equation}
\varepsilon_0 c n_1 (E_1^{+}e^{i k_{1} d_1} - E_1^{-}e^{-i k_{1} d_1}) = \varepsilon_0 c n_2 (E_2^{+}e^{i k_{2} d_1} - E_2^{-}e^{-i k_{2} d_1}) +  \sigma d(E_1^{+} e^{i k_{1} d_1} + E_1^{-} e^{-i k_{1} d_1})
\label{eq8}
\end{equation}
\end{widetext}
Finally, for the interface between the 2nd and 3rd zones, the continuity of the tangential component and zero value of the electric field inside the PEC reflector gives:
\begin{equation}
E_2^{+} e^{i k_{2} (d_1 + d_2)} + E_2^{-} e^{-i k_{2} (d_1 + d_2)} = 0
\label{eq9}
\end{equation}
Let us introduce notations $r$, $r_{123}$ and $y$, which are defined as:
\begin{equation}
E_0^{-} = r E_0^{+},
\label{eq10}
\end{equation}
\begin{equation}
E_1^{-} e^{-i k_{1} d_1} = r_{123} E_1^{+} e^{i k_{1} d_1}.
\label{eq11}
\end{equation}
\begin{equation}
y = \frac{\sigma d}{\varepsilon_0 c},
\label{eq12}
\end{equation}
Each of the three quantities defined by Eqs. (\ref{eq10})-(\ref{eq12}) have a physical meaning. $r$ is the amplitude reflection coefficient for the plane wave, incident on the structure. $r_{123}$ is the amplitude reflection coefficient for the 1-2 interface, taking into account the interference in the bottom lossless layer. Using the definitions of the thin film sheet resistance $R_{sh} = 1/\sigma d$ and of the vacuum impedance $Z_0 = \sqrt[]{\mu_0/\varepsilon_0}$ one can show that $y$ represents the ratio of $Z_0$ to $R_{sh}$:
\begin{equation}
y = \frac{Z_0}{R_{sh}},
\label{eq13}
\end{equation}
Putting the defined quantities $r$, $r_{123}$ and $y$ in the system of linear equations (\ref{eq5})-(\ref{eq9}) and solving yields the following expression for the amplitude reflection coefficient and electric field amplitudes in the 1st and 2nd zones:
\begin{equation}
r = \frac{r_{01} + r_{123} e^{2i k_1 d_1}}{1 + r_{01} r_{123} e^{2i k_1 d_1}},
\label{eq14}
\end{equation}
where
\begin{equation}
r_{01} = \frac{n_{0} - n_{1}}{n_{0} + n_{1}},
\label{eq15}
\end{equation}
and 
\begin{equation}
r_{123} = \frac{n_1 - n_2 - y + (y - n_1 - n_2) e^{2i k_{2} d_2}}{n_1 + n_2 + y + (n_2 - n_1 - y) e^{2i k_{2} d_2}}.
\label{eq16}
\end{equation}
Equations for the electric field amplitudes are:
\begin{equation}
E_1^+ = \frac{2 n_0 E_0^+}{n_0 + n_1 + (n_0 - n_1) r_{123} e^{2 i k_1 d_1}},
\label{eq17}
\end{equation}
\begin{equation}
E_2^+ = -\frac{e^{i k_1 d_1 - i k_2 d_2} E_1^+ (1 + r_{123})}{-1 + e^{2 i k_2 d_2}},
\label{eq18}
\end{equation}
\begin{equation}
E_2^- = \frac{e^{i k_1 d_1 + i k_2 (d_1 + 2 d_2)} E_1^+ (1 + r_{123})}{-1 + e^{2 i k_2 d_2}}.
\label{eq19}
\end{equation}
The amplitude $E_1^-$ can be calculated using Eq. (\ref{eq11}).

One can straightforwardly calculate the intensity reflection coefficient as
\begin{equation}
R = |r|^2,
\label{eq20}
\end{equation}
and the absorption coefficient as
\begin{equation}
A = 1-|r|^2.
\label{eq21}
\end{equation}
Note that Eqs. (\ref{eq14}-\ref{eq16}) and (\ref{eq21}) are already enough for the numerical analysis of absorption in the model structure. However, the absorption coefficient can also be calculated using two more alternative methods. These two additional methods are used in the next section to derive analytical equations describing the conditions of total absorption. First, the absorption can be calculated through the energy balance at the 1-2 interface, which states that the net power flux along $z$ axis in the 2nd zone equals net power flux along $z$ axis in the 1st zone minus ohmic losses at the 1-2 interface:

\begin{equation}
S_2^+ - S_2^- = S_1^+ - S_1^- - P_{A}
\label{eq22}
\end{equation}
where $S_j^+$ and $S_j^-$ are the Poynting vectors associated with the two counter-propagating plane waves in the $j$-th zone, and $P_{A}$ denotes the absorbed power per unit area of the film. Note that the assumption of PEC reflector results in $S_2^+ = S_2^-$. Hence, using the definition of the time-averaged Poynting vector one can write:
\begin{equation}
P_{A} = \frac{1}{2}\varepsilon_0 c n_1 (|E_1^+|^2 - |E_1^-|^2) = \frac{1}{2}\varepsilon_0 c n_1 |E_1^+|^2 (1 - |r_{123}|^2)
\label{eq23}
\end{equation}
Second, the absorption coefficient can be calculated using the Ohm’s law as:
\begin{equation}
P_{A} = \frac{1}{2} \sigma d |E_1^{+} e^{i k_{1} d_1} + E_1^{-} e^{-i k_{1} d_1}|^2  = \frac{1}{2} \sigma d |E_1^+|^2 |1+ r_{123}|^2.
\label{eq24}
\end{equation}
And finally, $P_{A}$ relates to the absorption coefficient as $A = P_{A}/S_0^+$.

\section{Conditions for total absorption}

In this section, we derive the optimal relationships between the structure parameters that result in total absorption or, equivalently, zero reflection at a given wavelength, $\lambda$.  For the sake of simplicity of analysis and representation of the results, normalized thicknesses for the lossless layers are defined as $\delta_1 = n_1 d_1/ \lambda$ and $\delta_2= n_2 d_2/ \lambda$. The condition for total absorption is obtained from the roots of the numerator of Eq. (\ref{eq14}):
\begin{equation}
r_{01} + r_{123} e^{4\pi i \delta_1} = 0,
\label{eq25}
\end{equation}
which can be rewritten in the form of two equations that must be simultaneously satisfied:
\begin{equation}
|r_{01}| = |r_{123}|,
\label{eq26}
\end{equation}
\begin{equation}
4\pi \delta_1 + \arg(r_{123}) = 2\pi m, m = 0,1,2,...
\label{eq27}
\end{equation}
It is important to note that neither $r_{01}$ nor $r_{123}$ depend on $\delta_1$. Hence Eq. (\ref{eq26}) can be satisfied for an arbitrary $\delta_1$ value. At the same time, Eq (\ref{eq27}) can also be satisfied for an arbitrary $\delta_1$ by the proper choice of $\delta_2$ and $y$. Hence, zero reflection can be achieved for an arbitrary $\delta_1$.

Eqs. (\ref{eq26}) and (\ref{eq27}) can be solved analytically in two special cases. The first solution is given by
\begin{subequations}
\begin{eqnarray}
y = n_0, 
\\
\delta_1 = k/2, 
\\
\delta_2 = (2m-1)/4, 
\end{eqnarray}
\label{eq28}
\end{subequations}
where $k=0,1,2,...$ and $m=1,2,3,...$ Note that the solution given by Eq. (\ref{eq28}) at $k=0$ and $m=1$ represents the standard quarter-wavelength absorber structure. The second solution is given by
\begin{subequations}
\begin{eqnarray}
y = n_1^2/n_0, 
\\
\delta_1 = (2k-1)/4, 
\\
\delta_2 = (2m-1)/4, 
\end{eqnarray}
\label{eq29}
\end{subequations}
where $k=1,2,3,...$ and $m=1,2,3,...$
Eqs. (\ref{eq26}) and (\ref{eq27}) cannot be straightforwardly solved analytically for the general case. However, the solution can be approached in an alternative way. For that let us equate the right-hand sides of Eqs. (\ref{eq23}) and (\ref{eq24}):
\begin{equation}
\frac{1}{2}\varepsilon_0 c n_1 |E_1^+|^2 (1 - |r_{123}|^2) = \frac{1}{2} \sigma d |E_1^+|^2 |1+ r_{123}|^2.
\label{eq30}
\end{equation}
From Eqs. (\ref{eq12}) and (\ref{eq30}), one can derive
\begin{equation}
y = n_1 \frac{(1-|r_{123}|^2)}{|1+r_{123}|^2}.
\label{eq31}
\end{equation}
In the general case, Eq. (\ref{eq28}) is a meaningless mathematical identity, which can be checked using Eq. (\ref{eq16}). And, obviously, this has to be so because $\lambda$, $d_1$, $d_2$, $n_0$, $n_1$, $n_2$ and $y$ are independently chosen. However, in the case of total absorption, the structure parameters cannot be independently chosen, and must satisfy Eq. (\ref{eq25}). Hence, one can rewrite Eq. (\ref{eq31}) in the following way 
\begin{equation}
y = n_1 \frac{(1-|r_{01}|^2)}{|1-r_{01} e^{-4\pi i \delta_1}|^2}.
\label{eq32}
\end{equation}
Eq. (\ref{eq32}) explicitly determines the optimal value of $y$ and, thus, the optimal relation between the conductivity and thickness of the thin absorber metal film for an arbitrary thickness of the top layer of the structure. Using Eqs. (\ref{eq13}) and (\ref{eq32}), one can derive the expression for the optimal sheet resistance of the absorbing metal film:
\begin{equation}
R_{sh} = \frac{Z_0}{n_1} \frac{|1-r_{01} e^{-4\pi i \delta_1}|^2}{(1-|r_{01}|^2)}
\label{eq33}
\end{equation}

\begin{figure}
\begin{subfigure}[b]{0.8\linewidth}
\hfill
    \centering
    \includegraphics[width=1\textwidth]{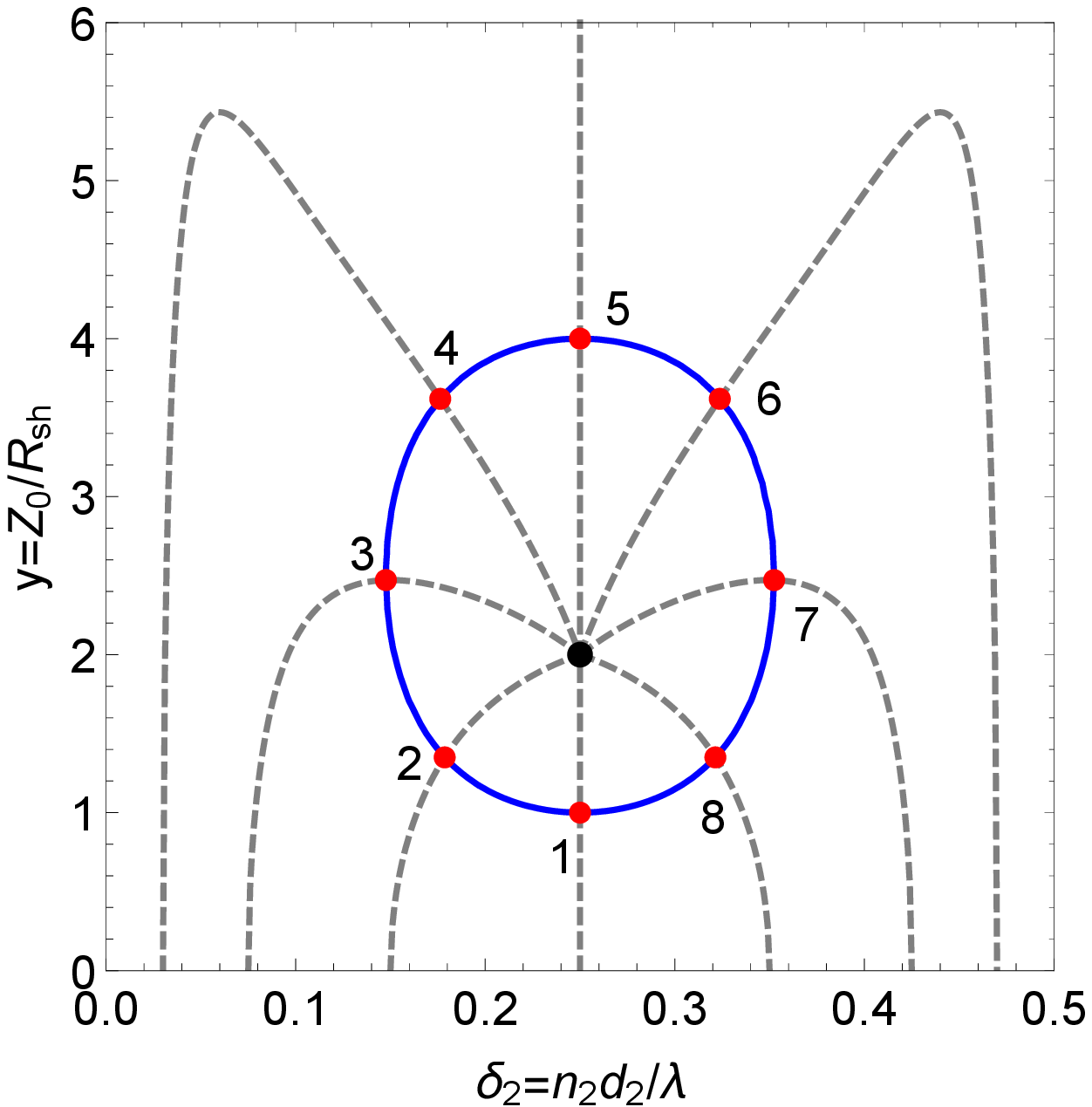}
    \caption{}
    \label{fig3a}
    \end{subfigure}
    \hfill
    \begin{subfigure}[b]{0.8\linewidth}
    \centering
    \includegraphics[width=1\textwidth]{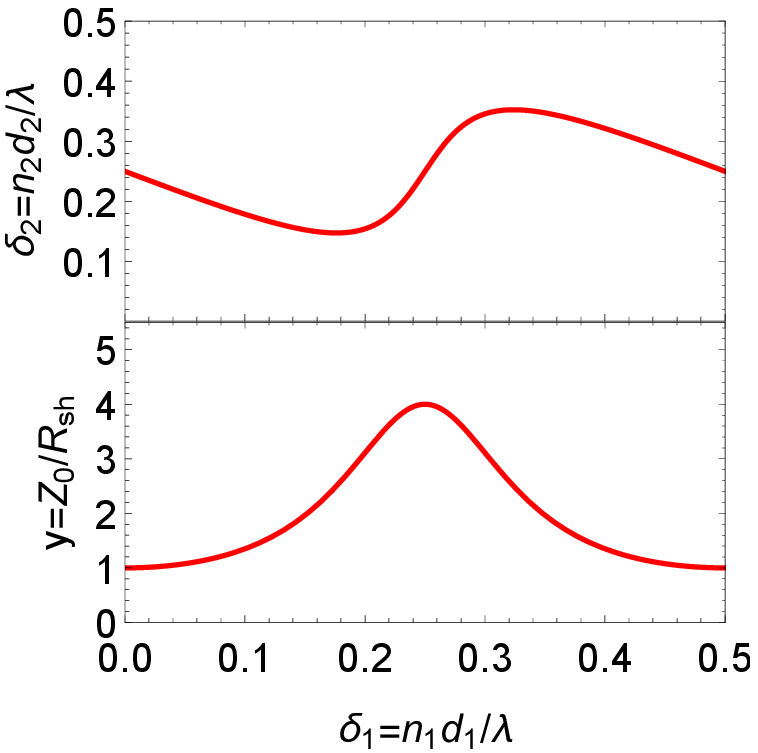}
    \caption{}
    \label{fig3b}
    \end{subfigure}
    \label{fig3}
    \caption{a) graphical illustration of solution of Eqs. (\ref{eq26}) and (\ref{eq27}) for $n_0=1.0$ and $n_1=n_2=2.0$. Blue contour represents solution for Eq. (\ref{eq26}); Grey dashed curves represent solution of Eq. (\ref{eq27}) for eight different values of $\delta_1$: 1) $\delta_1=0$, 2)  $\delta_1=0.1$, 3) $\delta_1=0.175$, 4) $\delta_1=0.22$, 5) $\delta_1=0.25$, 6) $\delta_1=0.28$, 7) $\delta_1=0.325$, 8) $\delta_1=0.4$. Red point ($\delta_{2}=0.25$, $y=2$) is the point where the value of $\arg(r_{123})$ involved in Eq. (\ref{eq27}) becomes indefinite since $r_{123}$ turns zero. b) Roots of Eqs. (\ref{eq26}) and (\ref{eq27}) obtained by numerical solution for $n_0=1.0$ and $n_1=n_2=2.0$.}
\end{figure}

Further, we graphically and numerically analyze the conditions of zero reflection (total absorption) expressed in Eqs. (\ref{eq26}) and (\ref{eq27}). We start with a simplified case: $n_1=n_2$. Fig. \ref{fig3a} graphically illustrates the solutions of these equations for $n_0=1.0$ and $n_1=n_2=2.0$. The blue dashed contour, or $r$-contour,  corresponds to the amplitude matching condition given by Eq. (\ref{eq26}). The black solid curves, or phase curves, correspond to the phase matching condition given by Eq. (\ref{eq27}), for eight different top layer thicknesses: 1) $\delta_1=0$, 2)  $\delta_1=0.1$, 3) $\delta_1=0.175$, 4) $\delta_1=0.22$, 5) $\delta_1=0.25$, 6) $\delta_1=0.28$, 7) $\delta_1=0.325$, 8) $\delta_1=0.4$. 
Each phase curve crosses the iso-reflectance contour at one point ($\delta_{2,0}$, $y_0$), corresponding to zero-reflection. For increasing $\delta_1$, the zero-reflectance point moves clockwise along the $r$-contour starting from the point ($\delta_{2,0}=0.25$, $y_0=0$) at $\delta_1=0$. Fig. \ref{fig3b} shows the dependence of  $\delta_{2}$ and $y_0$ on $\delta_1$ ($0\leq\delta_1\leq0.5$) at zero reflectance. It is seen that the optimal values of both $\delta_2$ and $y$ change non-monotonically for increasing $\delta_1$. The value of $\delta_2$ oscillates in a sawtooth form around its median value of 0.25. The value of $y$ first increases and reaches its maximum of 4.0 at $\delta_1=0.25$ and then decreases back to 1.0 at $\delta_1=0.5$. 
Note that the maximum value of $y$ in Fig. \ref{fig3b} corresponds to the analytical solution given by Eqs. (\ref{eq29}). And the analytical solution given by Eqs. (\ref{eq28}) describes points ($\delta_1=0$, $\delta_2=0.25$, $y=1$) and ($\delta_1=0.5$, $\delta_2=0.25$, $y=1$).

\begin{figure}[htbp]
	\centering
    \begin{subfigure}[b]{0.46\linewidth}
    \centering
    \includegraphics[width=1.1\textwidth]{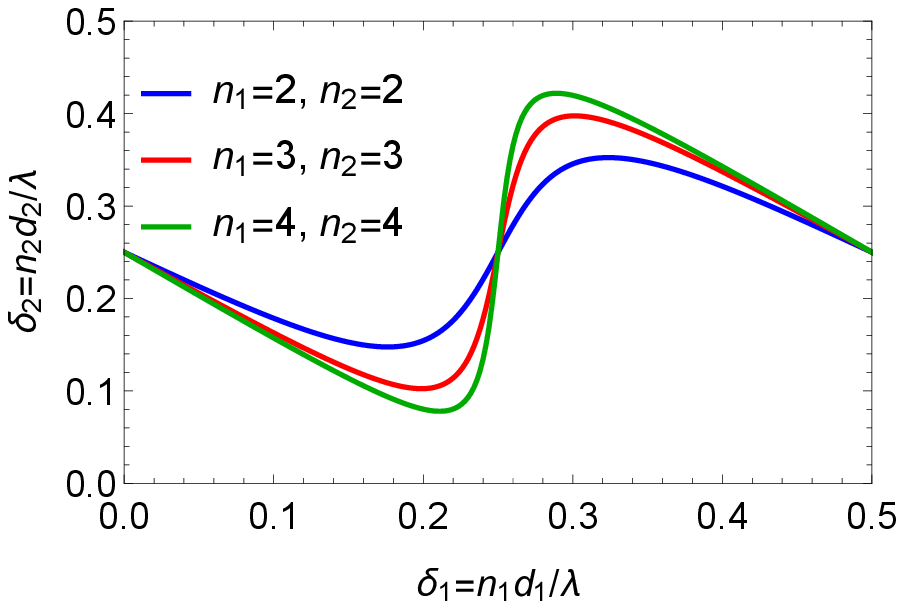}
    \caption{}
    \end{subfigure}
    \hfill
    \begin{subfigure}[b]{0.47\linewidth}
    \centering
    \includegraphics[width=1.1\textwidth]{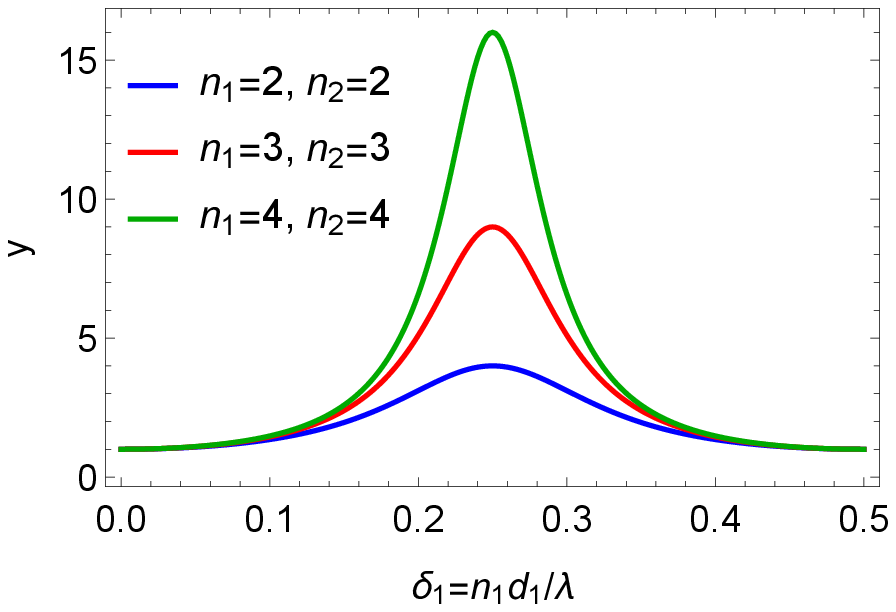}
    \caption{}
    \end{subfigure}
    \hfill
	\begin{subfigure}[b]{0.47\linewidth}
    \centering
    \includegraphics[width=1.1\textwidth]{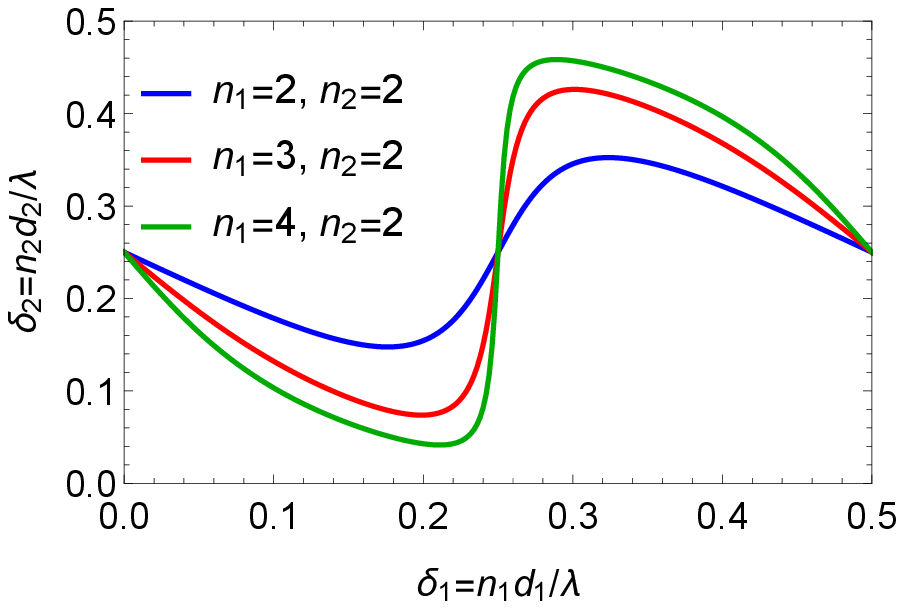}
    \caption{}
    \end{subfigure}
    \hfill
    \begin{subfigure}[b]{0.47\linewidth}
    \centering
    \includegraphics[width=1.1\textwidth]{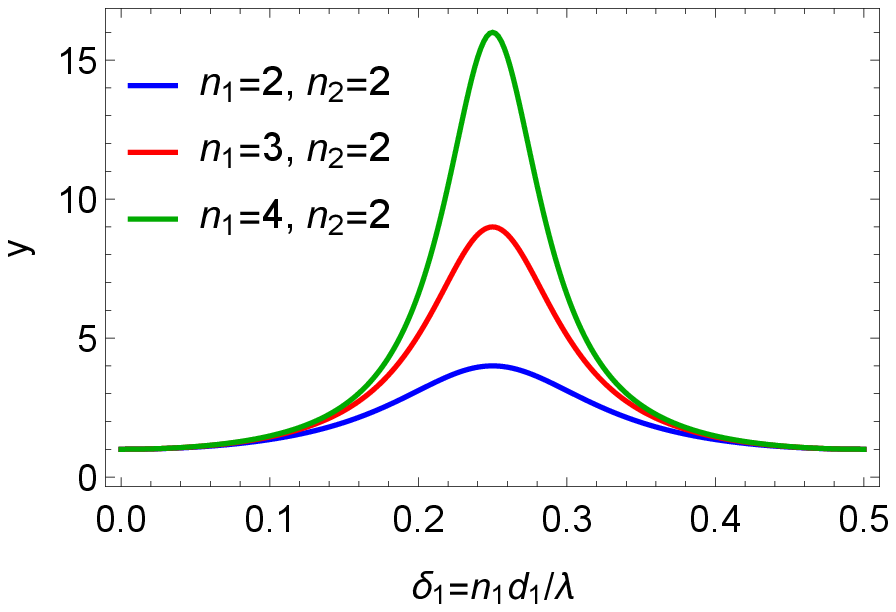}
    \caption{}
    \end{subfigure}
    \begin{subfigure}[b]{0.47\linewidth}
    \centering
    \includegraphics[width=1.1\textwidth]{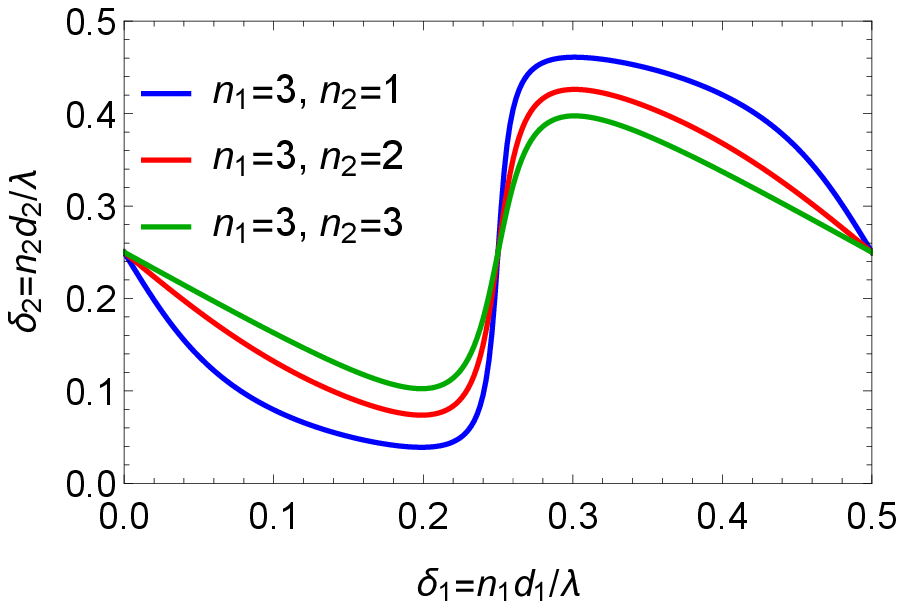}
    \caption{}
    \end{subfigure}
    \hfill
    \begin{subfigure}[b]{0.47\linewidth}
    \centering
    \includegraphics[width=1.1\textwidth]{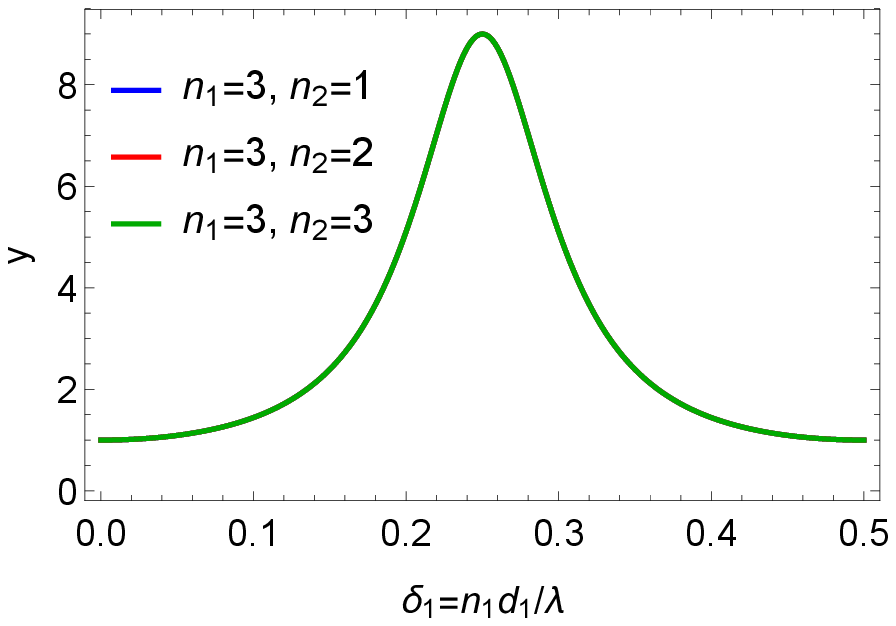}
    \caption{}
    \end{subfigure}
    \caption{a) and b) - calculated values of $\delta_2$ (a) and $y$ (b) providing total absorption: blue lines - $n_1=2, n_2=2$, red lines - $n_1=3, n_2=3$, green lines - $n_1=4, n_2=4$; c) and d) - calculated values of $\delta_2$ (c) and $y$ (d) providing total absorption: blue lines - $n_1=2, n_2=2$, red lines - $n_1=3, n_2=2$, green lines - $n_1=4, n_2=2$; e) and f) - calculated values of $\delta_2$ (e) and $y$ (f) providing total absorption: blue lines - $n_1=3, n_2=1$, red lines - $n_1=3, n_2=2$, green lines - $n_1=3, n_2=3$.}
    \label{fig4}
\end{figure}

Now we evaluate the effect of $n_1$ and $n_2$ variation on the optimal values of $\delta_2$ and $y$. First, we constrain the lossless layers to satisfy $n_1=n_2$. Figs. \ref{fig4}a and \ref{fig4}b shows the calculated optimal $\delta_2$ and $y$ values for three different values of the refractive index 2.0, 3.0 and 4.0. It is seen that the sawtooth profile of $\delta_2$ becomes sharper and the minimum (maximum) value of $\delta_2$ decreases (increases) with increasing refractive index. It is also noticeable that all three curves for $\delta_2$ cross each other at $\delta_1 = \delta_2 = 0.25$. The maximum value of $y$ grows in proportion to the square of $n_1$ as predicted by Eq. (\ref{eq29}) 
Figs. \ref{fig4}c and \ref{fig4}d describes the effect of $n_1$ variation at fixed $n_2$. It shows the calculated optimal $\delta_2$ and $y$ values for three different cases: blue lines - $n_1=2.0, n_2=2.0$, green lines - $n_1=3.0, n_2=2.0$, red lines - $n_1=4.0, n_2=2.0$. It can be seen that, at fixed $n_2$, the minimum (maximum) value of $\delta_2$ decreases (increases) for increasing $n_1$. Hence, for a given $n_2$, it is possible to reduce the spacing between the thin absorber film and reflector by increasing $n_1$. Figs. \ref{fig4}e and \ref{fig4}f describes the effect of $n_2$ variation at fixed $n_1$. It shows the calculated optimal $\delta_2$ and $y$ values for three different cases: blue lines - $n_1=3.0$, $n_2=1.0$, green lines - $n_1=3.0$, $n_2=2.0$, red lines - $n_1=3.0$, $n_2=3.0$. As expected from Eq. (\ref{eq32}), Fig. \ref{fig4}f shows that the optimal value of $y$ does not depend on $n_2$ at fixed $n_1$.

\section{Real materials}

\begin{table}
\caption{Calculated absorber thicknesses for various metals. The layer thickness in $d_{a}^*$ correspond to a standard QWA, while $d_{a}^{**}$ corresponds to the structure considered in this paper.}
\begin{ruledtabular}
\begin{tabular}{ccccc}
 metal              & $\sigma_0$ x 10$^6$ ($1/ \Omega m$) & $l_0$ (nm) & $d_{A}^*$ (nm) & $d_{A}^{**}$ (nm)\\
\hline
Ag & 63.0 & 53.3 & 0.93 & 4.02 \\
Cu & 59.6 & 39.9 & 0.84 & 3.64 \\
Au & 45.2 & 37.7 & 0.94 & 4.14 \\
Al & 37.7  & 18.9 & 0.74 & 3.44 \\
W  & 18.9  & 15.5 & 0.97 & 4.9 \\
Mo  & 18.7  & 11.2 & 0.84 & 4.42 \\
\end{tabular}
\label{tab1}
\end{ruledtabular}
\end{table}

There is a broad choice of infrared-transparent materials that can be used for the fabrication of dielectric layers. Fluorides, such as BaF$_2$ or YF$_3$, and CaF$_2$, possess low refractive indices, with $n < 2$. ZnS or ZnSe can be used as medium-$n$ materials, with $2 < n < 3$. Si and Ge can be used as high-$n$ materials, with $3 < n \leq 4$. As was mentioned in the introduction, metals that are typically used for the fabrication of the absorber layer in QWA structures are Ti, Cr and NiCr. Their relatively low electrical conductivities result in optimal absorber layer thicknesses in nanometer-scale range, while metals with high conductivity require sub-nanometer absorber layer thicknesses. However, the additional dielectric layer on top of the absorber metal layer allows the optimal thickness of the latter to be increased. As follows from Eq. (\ref{eq32}), the maximal absorber layer thickness is given by
\begin{equation}
d = \epsilon_0 c n_1^2/n_0 \sigma.
\label{eq34}
\end{equation}
This is achieved when the optical thicknesses of both dielectric layers equal quarter of the target wavelength. Note that the optimal absorber layer thickness for the standard QWA structure is $d = \epsilon_0 c/ \sigma$, thus, the factor $n_1^2/n_0$ suggests significant thickness increase. To estimate the layer thickness for real materials, it is important to take into account the thickness dependence of the conductivity of metal films in nanometer-scale range \cite{wissmann2007electrical}. 
To be more precise, electrical conductivity decreases with decreasing film thickness. This dependence can be taken into account using the Fuchs-Sondheimer formula for the electrical conductivity of thin films:
\begin{equation}
\sigma = \sigma_0 (1 + \frac{3}{8} (1-p)l_0/d)^{-1},
\label{eq35}
\end{equation}
where $\sigma_0$ is the conductivity of the bulk metal, $l_0$ is the electron mean free path in the bulk metal, and $p$ is the fraction of electrons specularly reflected at the film surfaces. For any given metal, Eqs. (\ref{eq34}) and (\ref{eq35}) allow the optimal absorber layer thickness to be estimated for the considered structure. Table 1 shows results of calculations for five metals. In the calculations we used $n_1 = 4$, which corresponds to germanium. Data for $\sigma_0$ and $l_0$ is taken from \citet{gall2016}. The optimal thickness for a QWA structure made from the same metal is added for comparison. From Table 1 it is seen that the considered three-layer structure suggests the optimal thickness of the absorber layer in the nanometer range which is feasible for fabrication unlike the result corresponding to the standard QWA structure.

\section{Discussion}

It is important to comment on the feasibility of production of the considered three-layer absorber structures. The successful fabrication of similar structures combining BaF$_2$ and NiCr layers and structures combing Al$_2$O$_3$ with Ti layers was demonstrated by Guo et al \cite{Guo:14}. The authors used the electron beam evaporation technique for the deposition of BaF$_2$, NiCr and Ti layers and the atomic layer deposition technique to grow Al$_2$O$_3$ layers. Peng et al. successfully fabricated similar structures combining Si$_3$N$_4$ with Ti layers \cite{Peng:16}. These authors used the combination of the plasma-enhanced chemical vapor deposition with the electron beam evaporation. Medvedev et al. used the magnetron sputtering technique for the fabrication of similar structures combining B$_4$C, Si and Mo layers \cite{medvedev2011infrared}. Thus the possibility of manufacturing the absorber structures considered in this paper is beyond doubt.

In the recent years metamaterials or metasurfaces became an alternative to the planar layered absorber structures. There is a broad class of metamaterials that is based on the concept of the planar QWA structure \cite{rhee2014metamaterial,watts2012metamaterial}. More precisely, part of the top metal layer material of the QWA structure can be removed according to a certain design pattern using the electron beam or optical lithography. Obviously, the pattering of the absorber metal layer can be also applied to the considered in this paper structure. Then the patterned metal layer can be coated with the top dielectric layer. The resulting metamaterial would have extra degrees of freedom for the optimization of its spectral properties and thus would be extremely interesting for the future studies.

Finally, it is important to comment on the validity of the described above theory and on the derived results. We assumed that the optical response of thin metal films is governed by its electric conductivity. And as a result, the derived optimal thickness of the absorber layer is directly determined by the metal conductivity. Our assumption works well for mid-infrared and longer wavelength radiation. When moving to the shorter wavelengths (near infrared and visible range), intraband processes become important for the radiation absorption in metals. The optimal absorber thicknesses for e.g. visible light significantly exceeds that for the long wavelengths. For instance, functional metal layers in planar absorbers for solar energy conversion can have thicknesses in 10-20 nm range \cite{Chirumamilla:16,Chen:16}.

\section{Conclusions}

Our analysis of a three-layer absorber structure shows several advantages over a two-layer QWA. Firstly, in the regime that we investigated, we show that $y$ can vary from 1 to $n_1^2$. However, $y$ is just the ratio of the vacuum impedance and the metal layer's sheet resistance. The sheet resistance, for ultrathin layers, depends on the intrinsic properties of the metal, and the layer thickness. Hence, with a three-layer structure, one can maximize $y$, allowing for thicker layers of poor conductors, easing fabrication constraints. Alternatively, one can also use metals with better conductivity at layer thicknesses that are still within the range of fabrication. Furthermore, the three layer structure may protect the metal layer from oxidation, increasing its lifetime. Finally, although we do not investigate the absorption spectrum in this paper, the additional freedom of choosing layer thickness may also allow some degree of spectral tailoring.

\end{document}